\documentclass[journal,onecolumn]{IEEEtran}

\usepackage{graphicx}
\usepackage{subfig}
\usepackage{multirow}
\usepackage{array}
\usepackage{epstopdf}
\usepackage{epsfig} 
\DeclareGraphicsExtensions{.pdf,.eps,.png,.jpg,.mps}
\usepackage{minibox}

\usepackage{amsmath} 
\usepackage{amssymb}  
\usepackage{algorithm}
\usepackage{cite}
\usepackage{footnote}
\usepackage{mathtools}
\allowdisplaybreaks
\usepackage{soul}
\usepackage{flushend}
\usepackage{color}
\usepackage{supertabular}
\usepackage{calc}  
\usepackage{enumitem}  
\usepackage[colorinlistoftodos,prependcaption,textsize=tiny]{todonotes}

\title{ Optimizing Communication and Computation for Multi-UAV Information 
	Gathering Applications}
\author{Mason Thammawichai,
	Sujit P. Baliyarasimhuni,
	Eric C. Kerrigan	and Jo\~{a}o B. Sousa%
	\thanks{ Mason Thammawichai is with the 
		Department of Aeronautics, Imperial College London, London SW7 2AZ, UK. m
		E-mail: m.thammawichai12@imperial.ac.uk}
	\thanks{ Sujit P.~Baliyarasimhuni is with the Department of 
		Electronics and Communications Engineering, Indraprastha Institute of 
		Information Technology, Delhi, New Delhi, 110020, India. Email: sujit@iiitd.ac.in}
	\thanks{ Eric C.~Kerrigan is with the Department of Electrical \& 
		Electronic Engineering  and the Department of Aeronautics, Imperial College 
		London, London SW7 2AZ, UK.Email: e.kerrigan@imperial.ac.uk}
	\thanks{Jo\~{a}o B. Sousa is with the Department of Electrical 
		and Computer Engineering, University of Porto, Portugal - 4200-465. Email: jtasso@fe.up.pt}}

\begin{document}
	\maketitle
\IEEEpeerreviewmaketitle

\begin{abstract}
Mobile agent networks, such as multi-UAV systems, are constrained by limited resources. In particular, limited energy affects system performance directly, such as system lifetime. It has been demonstrated  in the wireless sensor network literature that the communication energy consumption  dominates the computational and the sensing energy consumption. Hence, the lifetime of the multi-UAV systems can be extended significantly by 
optimizing the amount of communication data, at the expense of increasing computational
cost. In this work, we aim at attaining an optimal trade-off between the communication and the computational energy. Specifically, we propose a mixed-integer optimization formulation for a multi-hop hierarchical clustering-based self-organizing UAV network incorporating data aggregation, to obtain an energy-efficient information routing scheme. The proposed framework is tested on two applications, namely target tracking and area mapping. Based on simulation results, our method can significantly save energy compared to a baseline strategy, where there is no data aggregation and clustering scheme.
\end{abstract}


\section{Introduction}\label{sec:introduction}
Inexpensive mobile agents, such as unmanned aerial vehicles (UAVs), are  useful for several remote monitoring applications such as agriculture \cite{ROB:ROB20343}, geology \cite{nex2014}, ecology~\cite{anderson2013lightweight} and forestry \cite{wallace2012development}. The viability of UAVs for scientific and  non-military applications are due to 
reduced cost of the UAVs, low sensor cost and ease in handling. Typically, these applications are of large scale and the mission time can be  shortened by introducing multiple UAVs. 

Central to these applications is the necessity to have a human-in-the-loop (HITL) capability that increases situational awareness and  operator autonomy to modify missions dynamically. For HITL, UAVs have to gather and disseminate information  periodically to the operator who may be located at a distant (base station) from the operational arena. Typical information required at the base station is aerial footage~\cite{zhou09}, which is a communication intensive operation  consuming considerable energy. Unfortunately, low cost UAVs have  limited flight time due to battery/fuel capacity. Hence, there is a need to find different mechanisms by which flight time endurance can be increased. One way is to use gliders that take advantage of the updrafts to soar for long endurance~\cite{allen2005autonomous}. However, during soaring it is very  
difficult to maintain a good resolution of the terrain due to varying UAV height for mapping or surveillance applications. Instead, we propose to optimize the energy consumed by various units in a given aircraft to increase the flight time and hence the UAV team mission time.

For many applications \cite{ROB:ROB20343,wallace2012development}, it is necessary that a UAV must fly at a constant speed and maintain a prescribed height. Under these conditions, the major energy consumption units are propulsion, sensing, computation and communication. On average, the power consumed during flight is approximately constant. The sensing and the computational units also consume constant power.  However, the energy expended by the communication depends on (i) the amount of data to be transmitted, (ii) the distance between a vehicle and the base station and (iii) the number of vehicles transmitting data to the base station. Moreover, the communication cost is far greater than the sensing and computational energy. For example, a typical sensor node consumes 1\,nJ-1\,$\mu$J/sample, roughly 1\,pJ/instruction for computation, while communicating via radio frequency (RF) at the cost of 100\,nJ-50\,$\mu$J per bit~\cite{Doherty01}. Hence, it is better for the UAVs to cooperate with each other to minimize the team communication energy by performing computation on-board such that  the amount of data to be transmitted is  minimized. That is, optimally selecting (a) which vehicles should be the computing nodes and (b) determining how many vehicles are required to communicate with the base station. In this paper, we propose a general Mixed 
Integer Nonlinear Program (MINLP) that determines an optimal solution to (a) and (b).

\subsection{Related Work}
Similar to our Multi-UAV information gathering problem, the goal of a Wireless Sensor Network (WSN) is to maximize network lifetime while delivering raw data to the sink (base station)~\cite{yunxia2005}. In order to maximize the lifetime of a network, data aggregation techniques have been proposed for WSNs where some computations are performed within the node to reduce the communication cost. It has been shown that by using a sensor node as a communication relay/aggregator, an energy-efficient communication strategy can be obtained~\cite{bhard2002,yean2007}. Data correlations between different sensor nodes can be exploited to minimize the number of sensors sending the data to the base station~\cite{gupta08}. A compressed sensing technique to reduce the data volume to be transmitted was proposed in~\cite{xiang2013}. 

Hierarchical Network Routing is also one of the techniques in prolonging a network lifetime. For this approach, the nodes are grouped into clusters and  the cluster-head for each group is selected based on various election algorithms~\cite{nikolaos2013}. The cluster head is responsible for aggregation, compression and forwarding data to the base station. For example, in the Low-Energy Adaptive Clustering Hierarchy (LEACH) protocol proposed in~\cite{Hein2000}, a stochastic scheme is used to determine whether a node will become a cluster-head in each decision making round, i.e.~the probability that a node will become a cluster head is $1/P$, where $P$ is the desired percentage of cluster heads. The Low-Energy Adaptive Clustering Hierarchy Centralized (LEACH-C) protocol~\cite{Hein2002}, which is an improvement of LEACH, uses global information of the network to determine an optimal number of cluster heads via a centralized control at the base station. A chain-based protocol, called Power-Efficient Gathering in Sensor Information Systems (PEGASIS), where the nodes are only allowed to communicate with nearby nodes and take turns to transmit data to the base station, was proposed in~\cite{lindsey2002}. A hierarchical data aggregation technique where sensor nodes were grouped into clusters was proposed in~\cite{karaki2004}. A local aggregator (LA) for each cluster was selected, then a set of master aggregators (MAs) were selected based on LAs. To select  MAs, an integer program is solved such that the total communication energy is minimized, while performing minimum aggregation computation, such as finding an average or a maximum. For this work, we adopt a hierarchical cluster-based data aggregation technique from the WSN literature, but the topology of the network and the number of MAs are dynamically decided. 


Another approach is to have a mobile sensing node collect data from the nodes to reduce the communication overload~\cite{tekdas2009,sugihara2008,sujit2012,ho2013}. Since the UAVs are mobile, using another UAV to collect data from the surveying UAVs is not an ideal approach. However, similar to WSN data aggregation, the UAVs can perform computations on board to produce concise data and periodically transmit to the base station, as in~\cite{shum1999} for an image processing application. Data transmission to the base station can be performed either directly or through a UAV relay network~\cite{zhan2011,ponda2012}. Therefore, in this work, we propose a self-organizing network topology that allows data aggregation as well as a multi-hop information routing pattern.  

A UAV with sensing capabilities can be applied to perform target tracking due to its adaptability, scalability and better performance than a static wireless sensor network. However, most of the work on UAV target tracking applications only focus on the target tracking accuracy, while the communication and computation energy consumption has been neglected~\cite{pack2006,quin2010,wang2010,adurthi2014}. Hence, this work aims to incorporate both the communication and computation energy consumption into a multi-UAV target tracking application. Target tracking algorithms are based on target state estimation. By combining multiple sensor readings, which originated from different moments in time and distances from the UAVs, a more accurate state estimate can be obtained~\cite{kaland2004}. Precisely, the tracking objective is to maximize the information contribution~\cite{kalandros2002,kaland2004} from each node. In general, it has been shown that the measurement obtained from the most distant node does not contribute much to the target tracking accuracy. Therefore, it would be energy-efficient to select only the subset of the UAVs to be tracking nodes. The problem of deciding a subset of tracking sensor nodes could be formulated as an MINLP as in~\cite{ling2011}, where the observation covariance depends on the distance, i.e.~the further away from the target, the less accurate the measurement. Therefore, in this work, we include the information contribution constraint to our optimal control formulation for a target tracking application. 
 

 UAVs have been used for mapping applications \cite{eisenbeiss2004mini,nex2014uav,ROB:ROB20343,fornace2014mapping}. However, the focus of mapping applications using UAVs has been on improving the accuracy of the acquired images, which could be orthomosaic, classification of vegetation, improving  video transmission range, etc.  In some applications, the objective is to determine the minimum energy cost path for UAVs. In \cite{di2015energy}, the objective for the UAV is to visit a set of pre-defined target locations. The determined path must  minimize the total energy consumed in visiting the targets. In \cite{chung2015variable},  the objective is to develop multi-UAV exploration strategies under limited battery constraints.  In \cite{cesare2015multi}, a multi-UAV cooperative system using behavior was developed to efficiently explore a region with the constraint that the UAVs have limited energy. In most of the above UAV mapping applications, the issue of optimizing communication energy to enhance mission time is not considered. In our formulation, we want to optimize the energy consumed by  communication and computation components, so that the mission duration can be increased. This aspect has not be adequately addressed in the UAV mapping literature.
 
 \subsection{Contribution}
This paper proposes a simple optimal control problem for mobile agent systems with the objective of minimizing the communication and the computation energy. Particularly, we present an MINLP formulation for a multi-hop hierarchical cluster-based self-organizing UAV network to attain an energy-efficient reporting mechanism. The main contributions of this work are:
\begin{itemize}
\item A general MINLP optimization framework for a multi-UAV network to optimally trade-off between the communication and the computational energy was presented. That is, to dynamically determine: (i) the optimal number of agents to communicate to the base station, (ii) the role of each UAV: a sensor, a relay or an aggregator, (iii) the communication links among the UAVs to obtain an energy-efficient information routing network with data aggregation. 
\item Our data aggregation network model exploits three benefit characteristics: (i) a self-organizing network, which means that the topology of the network is dynamically decided at each decision time interval, resulting in a more flexible and reliable network, (ii) a multi-hop network, which exploits the shorter communication distance to prolong the lifetime of the network and (iii) a hierarchical clustering network, which can provide a better performance in terms of energy consumption, reliability, as well as scalability. 
\item A generalised data aggregation network model that allows multiple flows of more than one data type within the network. In other words, our network model can be applied to a heterogeneous mobile computing system, where only the same data types are allowed to be aggregated/processed, i.e.~a system with more than one sensor type. 
\item Two information gathering applications, namely target tracking and area mapping are addressed by our proposed optimal control framework to illustrate both the correctness and the effectiveness in trading off communication and computation energy.
\item Simulation results show an energy saving of up to 40\% for target tracking and 60\% for area mapping when comparing the performance of our MINLP formulation with a baseline approach, where there is no data aggregation and clustering scheme.
\end{itemize}

\subsection{Notations}
This section provides summary of all notations used throughout the paper.

\begin{description}[leftmargin=!,labelwidth=\widthof{longest}]
\item[Variables:]
\item[\textnormal{Symbol}]  Description  
\item[$N$]  Set of all UAVs (nodes)
\item[$n$]  Total number of nodes 
\item[$C$]  Communication link matrix/vector 
\item[$c$]  Communication link assignment 
\item[$M$]  Set of all data types 
\item[$m$]  Total number of data types 
\item[$\lambda$] Average data transmitting rate 
\item[$\overline{\lambda}$]  Sensing rate 
\item[$\epsilon$]  Sufficiently small constant/energy constant
\item[$|G|$]  Total number of sensors of a data type
\item[$a$]  Aggregator assignment
\item[$\gamma$] Aggregator ratio
\item[$B$]  Communication bandwidth
\item[$h$] Decision time interval length
\item[$E$] Energy consumption
\item[$d$] Distance between nodes
\item[$e$] Energy state vector
\item[$\phi$] Inertial position vector
\item[$x$] Position in x-axis
\item[$y$] Position in y-axis
\item[$v/V$] Speed/speed vector
\item[$\phi/\Phi$] Heading angle/Heading angle vector
\item[$r$]  Distance/range
\item[$X$] State of the system
\item[$u$] Control input
\item[$\pi$] Information contribution
\item[$H$] Observation matrix
\item[$R$] Measurement noise covariance matrix
\item[$F_0$] State transition matrix
\item[$w_0$] Process noise vector
\item[$Q_0$] Noise covariance matrix
\item[$Z$] Measurement vector
\item[$\nu$] Measurement noice vector
\item[$K$] Distance-independent coefficient
\item[$Q$] Information matrix
\item[$P$] Covariance error matrix
\item[$\hat{q}$] Information state vector
\item[$S$] Set of sensor nodes
\item[$W$] Width of a region
\item[$T$] Length of a region 
\item[$\zeta$] Overlap factor
\item[$N_\ell$] Total number of lanes
\item[$\ell$] Lane
\item[$\omega$] Waypoint 
\item[$\tau$] Transition boundary
\item[$\chi$] Entry angle
\end{description}

\begin{description}[leftmargin=!,labelwidth=\widthof{longest}]
\item[Subscritpts/Superscripts:]
\item[\textnormal{Symbol}] Description 
\item[$i,j$]  UAV (node)
\item[$0$] Source (target) node/initial state
\item[$n+1$]  Sink node (base station)
\item[$z$] Data type
\item[$c$] Communication
\item[$s$] Sensing
\item[$p$] Processing 
\item[$t$] Transmitting (section~\ref{comm})/top (section~\ref{mapp})
\item[$r$] Receiving
\item[$\beta$] Path loss exponent
\item[$+$] Next state
\item[$k$] Decision making round
\item[$\kappa$] Lane index
\item[$b$] Bottom
\item[$d$] Desired heading angle
\end{description}

\subsection{Outline of Paper}
The rest of the paper is organized as follows: In Section~\ref{sec:appDetails}, the application details are presented. Details on problem assumptions, system models and variable definitions are given in Section~\ref{sec:pform}. The optimal control problem formulation is presented in Section~\ref{sec:optimalProblem}. The optimal control problem is applied to 
target tracking and mapping applications in Section~\ref{sec:applications} as well as simulation results. We conclude in Section~\ref{sec:conclusions}.

\section{Application Details} \label{sec:appDetails}
For this project, we are looking at the scenerio where a team of $n$ UAVs is given a mission to either pursue a single target or survey an area of interest (AOI) and needs to periodically send the data back to the base station. 

\subsection{System Assumptions} 
We will assume that at each decision making time interval, each UAV (node) $i\in N:=\{1,\ldots,n\}$ has the same capability of sensing, data aggregation and communication functions, where $n$ is the total number of UAVs in the fleet. A UAV can reach any UAV using one-hop communication. A sensing UAV periodically senses a target/AOI, i.e.~information (a data packet) is generated at a constant rate,   and hence, the energy consumed by the sensor is constant. We assume that the UAVs are flying at constant altitude having constant speed and there are no wind disturbances. The power consumed by the propulsion unit during level flight is given by the relation  \cite{noth2008design},  \begin{eqnarray} P_{prop} = \frac{C_D}{C_L^{3/2}}\sqrt{\frac{2Rg^3}{\rho}} \frac{m^{\frac{3}{2}}}{b},\end{eqnarray} where $C_D$ is the drag coefficient, $C_L$ is the lift coefficient, $R$ is the aspect ration of the aircraft, $g$ is the gravity constant, $\rho$ is the air density, $m$ is the mass of the aircraft and $b$ is the wing span. As we can see from the relation that for a level flight, all the quantities associated with $P_{prop}$ are constant. Further, since we assume that the UAV is flying at a fixed altitude, the lift and drag cooefficients that depend on the velocity of the aircraft are also constant. Hence we assume that the energy consumed by the propulsion unit is constant under these assumptions.  Since  the sensor and propulsion energy consumption is constant, including this in the formulation does not affect the decision-making. Hence, we do not consider this in our formulation. The information can be of different types, therefore our model can be thought of as either a single source or multiple sources with different data types. For simplicity, we will consider a system with only one base station to report the data. Note that extension to multiple sink nodes (base stations) is relatively straightforward.

\subsection{UAV as a Mobile Computing Node}

\begin{figure}[t]
\centering
\includegraphics[width=0.6\textwidth]{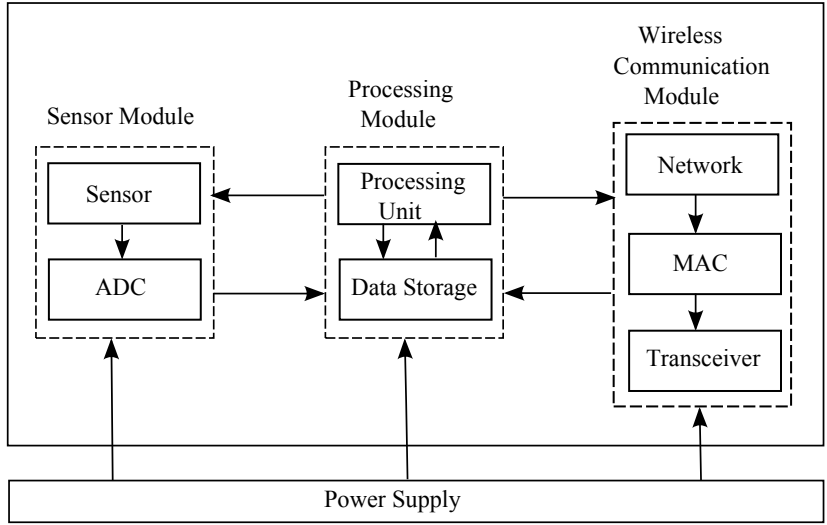}
\caption{The architecture of a mobile computing node (adapted 
from~\protect\cite{nikolaos2013})}
\label{fig:nodeArch}
\end{figure}

For this work, a UAV will be modeled as a mobile computing node, which is composed of three primary modules: a sensor module, a processing module and a wireless communication module, where interactions between modules are shown in Figure~\ref{fig:nodeArch}. The detailed description of each module is as follows:

\begin{itemize}
\item\textit{Sensor Module}: The main activities of this module includes sensing, analog to digital conversion (ADC) and signal modulation. 
\item\textit{Processing Module}: The processing module is responsible for data processing, sensor control as well as the communication protocol. 
\item\textit{Wireless Communication Module}: The wireless communication module is used for transmitting and receiving. We will assume that there exists a medium access control (MAC) protocol, which allows a UAV to communicate with other UAVs and the base station within a transmission range.
\end{itemize}  

\subsection{UAV Role Assignment}
Following the works of~\cite{bhard2002} and~\cite{patel2007}, we will assume that the UAVs can be assigned to one or more of the following roles at each time interval: (i) a sensor, which observes the target/AOI (called node $0$), via a sensor and produces the data which will be relayed to the base station (called node $n+1$), (ii) a relay, which simply relays its own data to the next level node without any processing, or (iii) an aggregator, which receives one or more data from other nodes, then aggregates the data of the same type to produce a single data point and sends the aggregated data to the next level node. 

\begin{figure}[t]
\centering
\includegraphics[width=0.45\textwidth]{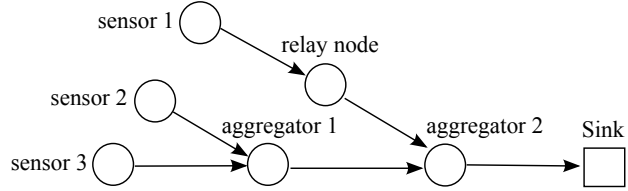}
\caption{Information Flow in an Aggregation Network Topology} \label{network}
\end{figure}

\subsection{Aggregation Network Topology}
Figure~\ref{network} illustrates the information flow in an aggregation network topology. In particular, the data obtained from the source (target/AOI) can be processed within the aggregator or passed along the relay node and routed to the sink (base station). Note that, in this work, the network topology is dynamic, which differs from others in the WSN literature, i.e.~the roles of the UAVs are decided at each time interval. Moreover, only data of the same type is allowed 
to be compressed/aggregated.

\section{Dynamic Model with Constraints} \label{sec:pform}

\subsection{Communication Model and Constraints}\label{comm}
Let $C:=[c_{ijz}]$ denote a communication link matrix, i.e. $c_{ijz} = 1$ if node $i$ transmits data of type $z$ to node $j$ for $i,j \in N^+:=N \cup \{0,n+1\},~z\in M:=\{1,\ldots,m\}$. Note that $c_{0iz} = 1$ if node $i$ is a sensor of data type $z$ and $c_{i(n+1)z} = 1$ if node~$i$ sends data type $z$  to the base station. The communication link matrix $C$ is subject to  
	\begin{align} \label{c}
	&c_{ijz}\in \{0,1\},&&\forall i\in N^+,j\in N^+,z\in M\\
	&\sum_{j=1}^n c_{0jz} \geq 1,&&\forall z\in M \label{c1}\\
	&\sum_{i=1}^n c_{i(n+1)z} \geq 1,&&\forall z\in M \label{c2}\\
	&\sum_{j=1}^{n+1} c_{ijz} \leq 1,&&\forall i\in N,z\in M\label{c3} \\
	&c_{iiz} = 0,&&\forall i\in N^+,z\in M \label{c4}
	\end{align}
where~(\ref{c1})--(\ref{c2}) guarantee that for each information type there is at least one communication link from a source to a node and there must be at least one communication link between a node and the base station, respectively. Note that contraint~\eqref{c1} defines an initial state of the network flow at each decision time interval. Constraint~(\ref{c3}) enforces that there is only one communication link of each data type out of a node. Constraint~(\ref{c4}) prevents self communication.

Let $\lambda_{ijz} \geq 0$ denote the average rate (packets per second) at which data of type $z$ is transmitted from node $i$ to node $j$. Note that $\lambda_{0jz}$ represents the sensing rate of data type $z$, assumed to be a constant equal to 
$\overline{\lambda}_z$ packets per time interval. Following the definition of the communication link matrix $C$, $\lambda_{ijz}$ needs to satisfy:
\begin{subequations}\label{c6}
\begin{align}
	&\lambda_{ijz} = 0 \Rightarrow c_{ijz} = 0,~\forall i\in N^+,j\in N^+,z\in M, \\
	&\lambda_{ijz} > 0 \Rightarrow c_{ijz} = 1,~\forall i\in N^+,j\in N^+,z\in M.
	\end{align}
\end{subequations}
Constraint~(\ref{c6}) says that if there is data flow between two nodes, then the link assignment should be active. The constraint~(\ref{c6}) can be implemented as the following inequality constraints:
\begin{equation}
\epsilon c_{ijz}\leq \lambda_{ijz} \leq |G_z|\overline{\lambda}_z c_{ijz},~\forall i\in N^+,j\in N^+,z\in M, \label{c7}
\end{equation}
where $\epsilon$ is a sufficiently small positive number and $|G_z|$ is the total number of sensors of data type $z$. In other words, suppose $\lambda_{ijz} \neq 0$, then~(\ref{c7}) is true if and only if $c_{ijz} = 1$. Suppose 
$\lambda_{ijz} = 0$, then~(\ref{c7}) is true if and only if $c_{ijz} = 0$.

Denote $a_{iz} \in \{0,1\}, \forall i\in N,~z\in M$ as the data type aggregator assignment, where by definition
\begin{equation}
a_{iz} = 1 \iff \sum_{j=0}^n c_{jiz} > 1,~\forall i\in N,z\in M. \label{c8} 
\end{equation}
In other words, if there are more than one packets of the same data type transmitted to a node, then the node will act as an aggregator. Constraint~(\ref{c8}) can be written as a set of linear inequalities as follows:
\begin{subequations} \label{ai}
	\begin{align}
	&(1-n)a_{iz} + \sum_{j=0}^n c_{jiz} \leq 1,&&\forall i\in N,z\in M, \\
	&(1+\epsilon)a_{iz}-\sum_{j=0}^n c_{jiz} \leq 0,&&\forall i\in N,z\in M,
\end{align}
\end{subequations}
where $\epsilon$ is a sufficiently small positive number.To guarantee a feasible communication link, the data flow within the node needs to be conserved, i.e.~the incoming data equals the aggregated outgoing data:
\begin{equation}\label{3a}
\begin{aligned}
	&\sum_{z=1}^m\sum_{j=1}^{n+1}c_{ijz}\lambda_{ijz} = \sum_{z=1}^m\sum_{j=0}^n c_{jiz}\lambda_{jiz}(1+(\gamma_z-1)a_{iz}),\\
	&\forall i\in N,z\in M,
	\end{aligned}
\end{equation}
where $0\leq\gamma_z\leq 1$ is the  aggregation ratio of data type $z$. Observe that when $\gamma_z = 1$, then there is no data aggregation/processing. 

Since the nodes are communicating via wireless network, the channel bandwidth are shared among the nodes. This implies that communication between two nodes restrains available bandwidth to other neighbor nodes. Therefore, bandwidth limitation should be considered in our formulation as well, i.e.~all communication data (number of transmitting/receiving bits) should not be greater than the channel bandwidth limitation. Specifically, the bandwidth constraints can be formulated as
\begin{equation}\label{bw}
  \sum_{z=1}^m\sum_{j=1}^{n+1}c_{ijz}\lambda_{ijz}L + \sum_{z=1}^m\sum_{j=1}^n c_{jiz}\lambda_{jiz}L\leq Bh,~\forall i\in N,
\end{equation}
where $B$ is the channel bandwidth (bits per second), $h$ is the decision time interval and $L$ is the packet length (number of bits per packet). 

\begin{figure}[t]
\centering
\includegraphics[width=0.48\textwidth]{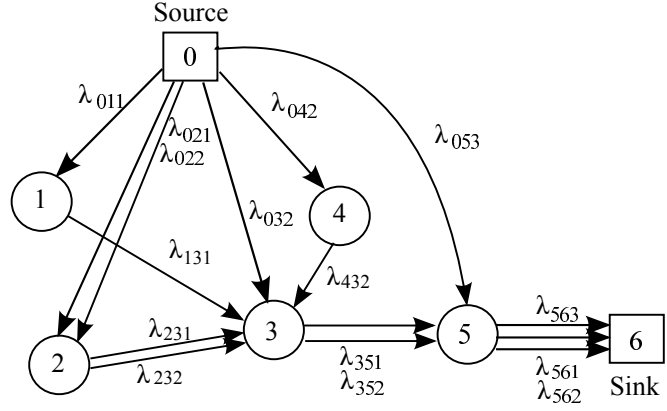}
\caption{Aggregation Network Topology Example} \label{ex1}
\end{figure}

Finally, we will use an example scenario to show the information flow topology that can be achieved from our model. Consider Figure~\ref{ex1} where the system is composed of five UAVs that are given a mission to retrieve three different types of information. Nodes 1,~2 and~4 are sensor nodes, node 3 is both a sensor and an aggregator, while node~5 is a sensor as well as a relay node. The correlated data obtained from node 1 $(\lambda_{131}$) and node 2 $(\lambda_{231})$ are processed within node 3. At the same time, the data obtained from nodes 2 $(\lambda_{232}),~3~(\lambda_{032})$ and 4 $(\lambda_{432})$ are also processed within node 3. Specifically, from ~(\ref{3a}), the outgoing data flow after the aggregation within node 3: $\lambda_{351}=(\lambda_{131}+\lambda_{231})\gamma$ and $\lambda_{352}=(\lambda_{032}+\lambda_{232}+\lambda_{432})\gamma$. Both processed data streams/packets are relayed to node 5, which are transmitted to the base station. Note that node 5 acts as a relay node because the data received from node 3 and its own data are of different types.

\subsection{Energy Models}
We will adopt an energy consumption model, which has been commonly used in the wireless sensor network literature~\cite{jin2003,yuan2006,bin2008}. The total energy in most multi-UAV applications is composed of three terms. The first term is the sensing energy $E^s$, which is the energy used to sense a target/AOI. We will assume that the energy to sense one bit of information is a constant equal to~$\epsilon_s$~J. The sensing energy consumed  by node $i$ within the time interval is
\begin{equation}
	E^s_i(c_{0iz}):=\epsilon_sL\sum_{z=1}^m\overline{\lambda}_zc_{0iz},~\forall i\in N.
\end{equation}
The second one is the aggregation energy $E^p$, which is the energy to do data processing. The energy to process one bit of information is also assumed to be a constant equal to $\epsilon_p$~J. The aggregation energy consumed by node $i$ within the time interval is
\begin{equation}
\begin{aligned}
E^p_i(c_{jiz},\lambda_{jiz},a_{iz}):=&\epsilon_pL\sum_{z=1}^m\overline{\lambda}_zc_{0iz}a_{iz}+ \\
	&\epsilon_pL\sum_{z=1}^m\sum_{j=1}^nc_{jiz}\lambda_{jiz}a_{iz},~\forall i\in N. \label{o3}
	\end{aligned}
\end{equation}
The last energy term is the communication cost, which is composed of two parts: the transmitting energy $E^t$ and the receiving energy $E^r$. The transmitting energy depends on the distance between the nodes $d_{ij}$, i.e.~$E^t(d_{ij}):=\epsilon_t+\epsilon_{rf}d_{ij}^\beta$, where $\beta \geq 2$ is the path loss exponent, $\epsilon_t$ (J/bit) and~$\epsilon_{rf}$~(J/bit/m$^\beta$) are constants. The energy of receiving one bit of information is assumed to be a constant equal to~$\epsilon_r$ J. The receiving energy consumed by node $i$ within the time interval is
\begin{equation}
	E^r_i(c_{jiz},\lambda_{jiz}):= 
\epsilon_rL\sum_{z=1}^m\sum_{j=1}^nc_{jiz}\lambda_{jiz},~\forall i\in N. 
\label{o4}
\end{equation}
 The transmitting energy consumed by node $i$ within the time interval is
\begin{equation}
  E^t_i(c_{ijz},\lambda_{ijz},d_{ij}):=\sum_{z=1}^m\sum_{j=1}^{n+1} 
(\epsilon_t+\epsilon_{rf}d_{ij}^\beta)c_{ijz}\lambda_{ijz}L,~\forall i\in N. 
\label{o5}
\end{equation}
The total energy used by node $i$ for sensing a target/AOI, processing information and communication during the time interval is denoted by
\begin{equation}
E_i := E^s_i+E^p_i+E^r_i+E^t_i,~\forall i\in N. \label{tenergy}
\end{equation}

Let $e_i$ be the energy stored in the $i^{th}$ UAV at time $t$, then the remaining energy $e_i^+$ at time $t+h$ is given by
\begin{equation}
e_i^+:=e_i-E_i \geq 0,~\forall i\in N. \label{eupdate}
\end{equation}  

\subsection{UAV Dynamic Constraints} \label{sec:dynconst}
The two-dimensional UAV kinetic model is given by:
\begin{equation} \label{uavdyn}
\begin{bmatrix}
\dot{x}_i \\ \dot{y}_i \end{bmatrix} =
f(\varphi_i,v_i,\psi_i) =
\begin{bmatrix}
v_i\cos\psi_i \\ v_i\sin\psi_i
\end{bmatrix},~\forall i\in N,
\end{equation}
where $\varphi_i = [x_i\ y_i]^T$ is the inertial position, $v_i$ is the speed and $\psi_i$ is the heading of the $i^{th}$ UAV. We will assume that UAVs fly at a constant speed and heading in the interval $[t,t+h]$ and are subject to the 
following constraint:
\begin{equation}\label{yawcons}
v_{\min} \leq v_i \leq v_{\max},~\forall i\in N,
\end{equation}
 where $v_{\min}$ and $v_{\max}$ are lower and upper bounds on speed. 

Moreover, since we assume that the UAVs are in one-hop communication range and to avoid collision among UAVs at each time interval, the following constraints are necessary:
\begin{equation} \label{colcons}
	r_c > d_{ij} \geq r_{safe},~\forall i\neq j, (i,j)\in N\times N,
\end{equation}
where $r_c$ is a sufficiently large positive number defined as a communication range limit, $d_{ij}$ is the distance between two nodes and $r_{safe}$ is the safety distance.
  
\subsection{State Update Equation}
Let $k$ denote the $k^{th}$ decision making round at time interval 
$[t_k,t_{k+1}]$, i.e. $t_{k+1}-t_k=h$. The state $X_i$ and the control input 
$u_i$ for the $i^{th}$ UAV are defined as
\begin{align}
&X_i:=(e_i,\varphi_i),&&\forall i\in N, \\
&u_{ijz}:=(c_{0iz},c_{ijz},\lambda_{ijz},a_{iz},v_i,\psi_i),&&\forall i\in N, z\in M,\nonumber\\
&&& j\in 
N\cup\{n+1\},
\end{align}
 where $X:=(X_1,X_2,\ldots,X_n)$ is the state of the overall system. The components of the overall system control input $u$ are all $u_{ijz}$, $ i\in N,j\in 
N\cup\{n+1\}, z\in M$.

Obviously, all the variables in the previous sections can be considered as a function of $k$. Let $X(k)$ denote the state 
of the overall system and $u(k)$ denote the system control input at time $t_k$. The overall system state update equation is given by
\begin{equation} \label{sdyn}
X(k+1) = \phi(X(k),u(k),k),~\forall k,
\end{equation}
 where $\phi$ can be derived from~(\ref{eupdate}) and~(\ref{uavdyn}). 

\section{Optimal Control Problem} \label{sec:optimalProblem}
We formulate the optimal control problem to determine the roles of the UAVs as an MINLP. We apply this formulation to a multi-UAV target tracking application and a multi-UAV mapping application. The MINLP is solved at each time instant $t_k$.

I) Target Tracking: Though our main objective is to minimize the total energy consumed by all nodes in the system~(\ref{tenergy}), for the target tracking application the target tracking accuracy should be considered as well. Particularly, this can be incorporated as a constraint that guarantees a minimum information contribution 
$\pi_{\min}$ requirement as
\begin{equation} \label{infocon}
	\pi:=\sum_{z=1}^m\sum_{i=1}^n c_{0iz} 
\operatorname{tr}\{H_{i}(t)^T\log{(R_{i}^{-1}(t))}H_{i}(t)\} \geq \pi_{\min},
\end{equation}
 where $\pi$ is the information contribution, $H_{i}(t)$ is the observation model and $R_i(t)$ is the measurement noise covariance. Note that our definition of information contribution is slightly different from the one defined and used in~\cite{kalandros2002,kaland2004,ling2011}. Specifically, we took the natural logarithm of the inverse of $R_i(t)$ to reduce the decay rate of information contribution in order to match with the target tracking application using mobile agents, i.e.~the useful information can be obtained within a reasonable distance between the sensor and the target.  

The sensing range limit can be implemented as the following constraint:
\begin{equation}\label{srange}
c_{0jz}(d_{0j}^2-r_{s}^2) \leq 0,~\forall j\in N,z\in M,
\end{equation}
where $d_{0j}$ is the distance between the node and the target and $r_{s}$ is the maximum sensing range. Constraint~(\ref{srange}) states that if a node is a sensor, then the distance between the node and the target has to be less than the maximum sensing range. Note that the square of the distance is chosen for an easier implementation.

The multi-UAV target tracking problem can be formulated as the following optimal control problem:
Given $n$ UAVs, a target and a base station, determine a role for each UAV, a communication network link and a UAV trajectory that solves
\begin{align}\label{prob}
&\underset{\begin{subarray}{c}
           u
\end{subarray}}{\text{minimize}}&&\quad\sum_{i=1}^n 
E_i \nonumber\\
	&\text{subject 
to}&&\text{(\ref{c})--(\ref{c4}),~(\ref{c7}),~(\ref{ai})--(\ref{colcons}),}\nonumber\\
	&&& \text{(\ref{infocon}) and~(\ref{srange})}\nonumber
\end{align}

II) Area Mapping: Given $n$ UAVs, an AOI, a base station and a UAV trajectory, 
determine a role for each UAV and a communication network link that solves
\begin{subequations}
\begin{align}
&\underset{\begin{subarray}{c}
          u\end{subarray}}{\text{minimize}}&&\quad\sum_{i=1}^n 
E_i \nonumber\\
&\text{subject 
to}&&\text{(\ref{c})--(\ref{c4}),~(\ref{c7}),~(\ref{ai})--(\ref{eupdate})~\text{and}}\nonumber\\
		&&& v_i = V_i, \forall i\in N,\\
		&&& \psi_i = \Psi_i^d, \forall i\in N,\\
		&&& c_{0iz} = C_i, \forall i\in N, \label{sensorAss}
\end{align}
\end{subequations}
where $V_i$ is the constant speed of the vehicle and $\Psi_i^d$ is the desired heading angle of the path. $C_i$ is a pre-determined data type sensor assignment vector. Also, note that for an area surveying/mapping application, the UAV dynamic constraints described in Section~\ref{sec:dynconst} are not included because we assume that the trajectory of each UAV and the collision avoidance among UAVs are decided by a path planning controller.

\section{Applications}\label{sec:applications}
This section provides simulation results to illustrate the correctness and effectiveness of our framework in trading off communication and computation energy consumption in multi-UAV applications. A multiple UAV single-target tracking and area mapping application are chosen as our demonstration examples. All simulations were simulated on MATLAB~\cite{MATLAB2012} and the MINLP was modelled using OPTI TOOLBOX~\cite{opti2012} and solved with SCIP~\cite{scip2009}.

\subsection{Target Tracking}
\subsubsection{Target and Sensor Models}
For a target tracking application, we will follow the work of~\cite{ling2011} to set up the optimization problem to make a decision on a subset of the UAVs to be sensor nodes. The motion of a target will be modelled as a linear discrete-time 
Markov process: 
\begin{equation} \label{tmodel}
X_0(t+1) = F_0(t)X_0(t) + w_0(t),
\end{equation}
where $X_0(t)$ is the state vector of a target, $F_0(t)$ is the state 
transition 
matrix and $w_0(t)$ is the process noise assumed to be zero mean Gaussian noise 
with covariance $Q_0(t)$.

The measurement equation of a sensor is
\begin{equation} \label{smodel}
Z_{i}(t) = H_{i}(t)X_0(t)+\nu_{i}(t),
\end{equation}
 where $\nu_{i}(t)$ is the measurement noise assumed to be zero mean Gaussian with covariance $R_{i}(t)$. We will assume that the measurement noise covariance is a function of the distance between a sensor and a target, i.e.~$R_{i}(t):=K(t)d_{0i}^\beta(t)$, where $K(t)$ is a distance-independent coefficient, and $d_{0i}(t)$ is the 
distance from a sensor to a target. Moreover, we will also assume that the measurement noise covariances are uncorrelated between any two nodes.

\subsubsection{Information Filter}
For multi-sensor data fusion, we use an information filter~\cite{vercau2005,ling2011}, which is an inverse covariance form of the Kalman filter. Let $\hat{X}_0(t|t)$ and $\hat{X}_0(t+1|t)$ denote the target estimated state vector and target predicted state vector, respectively. Define the information matrix $Q(t|t):=P^{-1}(t|t)$ and $Q(t+1|t):=P^{-1}(t+1|t)$, the information state vector $\hat{q}(t|t):=P^{-1}(t|t)\hat{X}_0(t|t)$ and $\hat{q}(t+1|t):=P^{-1}(t+1|t)\hat{X}_0(t+1|t)$, where $P(t|t)$ and $P(t+1|t)$ are the covariances of the estimation error $X_0(t|t)-\hat{X}_0(t|t)$ and the prediction error $X_0(t+1|t)-\hat{X}_0(t+1|t)$. The prediction and estimation steps are
\begin{align}
&\text{Estimation:}\nonumber\\
&\hat{q}(t|t)=\hat{q}(t|t-1)+H_i^T(t)R_i^{-1}(t)Z_i(t), \label{es1}\\
&Q(t|t)=Q(t|t-1)+H_i^T(t)R_i^{-1}(t)H_i(t), \label{es2}\\
&\text{Prediction:}\nonumber\\
&\hat{q}(t+1|t) = Q(t+1|t)F_0(t+1)Q^{-1}(t|t)\hat{q}(t|t),\\
&Q(t+1|t) = (F_0(t+1)Q^{-1}(t|t)F_0^T(t+1)+Q_0(t+1))^{-1}.
\end{align}

For multi-sensor data fusion, i.e~more than one node tracking the target,~(\ref{es1}) and~(\ref{es2}) are replaced, respectively by
\begin{align}
&\hat{q}(t|t)=\hat{q}(t|t-1)+\sum_{i\in S} H_i^T(t)R_i^{-1}(t)Z_i(t), 
\label{es1n}\\
&Q(t|t)=Q(t|t-1)+\sum_{i\in S} H_i^T(t)R_i^{-1}(t)H_i(t), \label{es2n}
\end{align}
where $S$ is a set of sensor nodes. 

\subsubsection{Simulation settings}\label{subsec:simSetting}For simplicity, we consider a small UAV network, i.e.~$n = 3$, which are deployed to track a single target in a two-dimensional area and needs to periodically report the target state back to the base station. Note that we consider the single target state as one data type. The base station is at 
(0,0). The initial positions of the UAVs are at positions (0,100), (100,0), and (100,100). The target initial position is (20,20). The target state vector $X_0(t)$ in~(\ref{tmodel}) is composed of the target positions in the $x$ and $y$ axes, and velocities in the $x$ and $y$ axes, denoted as $v_x$ and $v_y$, respectively. The parameters corresponding to the target state~(\ref{tmodel}), measurement equations~(\ref{smodel}) and information filter are~\cite{ling2011}:
\begin{equation}
F_0(t) =\begin{pmatrix}
1 & 0 & 1 & 0 \\
0 & 1 & 0 & 1 \\
0 & 0 & 1 & 0 \\
0 & 0 & 0 & 1
\end{pmatrix},~
Q_0(t) =\begin{pmatrix}
2 & 0 & 0 & 0 \\
0 & 2 & 0 & 0 \\
0 & 0 & 0.04 & 0 \\
0 & 0 & 0 & 0.04
\end{pmatrix},~\forall t\nonumber
\end{equation} 
\begin{equation}
H_i(t)=\begin{pmatrix}
1 & 0 & 0 & 0 \\
0 & 1 & 0 & 0
\end{pmatrix},~
K(t) =\begin{pmatrix}
1\times10^{-6} & 0\\
0 & 1\times10^{-6}
\end{pmatrix},~\forall t \nonumber
\end{equation}
\begin{equation}
\hat{q}(1|0)=\begin{pmatrix}
0 \\ 0 \\ 0 \\ 0 \end{pmatrix},~
Q(1|0) =\begin{pmatrix}
1 & 0 & 0 & 0 \\
0 & 1 & 0 & 0 \\
0 & 0 & 1 & 0 \\
0 & 0 & 0 & 1. \nonumber
\end{pmatrix}
\end{equation}

For all simulations, we let the target velocities be $v_x = 10$~\,m/s $v_y = 15$\,m/s. The UAV parameters~\cite{Kim2014} are $v_{\min} = 10$~\,m/s, $v_{\max} = 30$\,m/s, the initial UAV energy budget is $10$\,J, the communication range $r_c=500$\,m, the sensing range $r_{s}=200$\,m, the safety distance $r_{safe} =50$\,m, the decision time interval $h$ is 1\,s. The energy parameters~\cite{bhard2002} are $\epsilon_s =~50$\,nJ/bit,~$\epsilon_p =~10$\,nJ/bit,~$\epsilon_r =~135$\,nJ/bit,~$\epsilon_t=45$\,nJ/bit,~$\epsilon_{rf}=0.1$~\,nJ/bit/m$^2$,
~$\gamma_z = 0.7$, $\beta = 2$, $L = 1024$ bits/packet, $\overline{\lambda}_z = 5$ packets/time interval and $\pi_{min} = 6$. 

\subsubsection{Simulation Results}
\begin{figure}[t]
\centering
\includegraphics[width=0.5\textwidth]{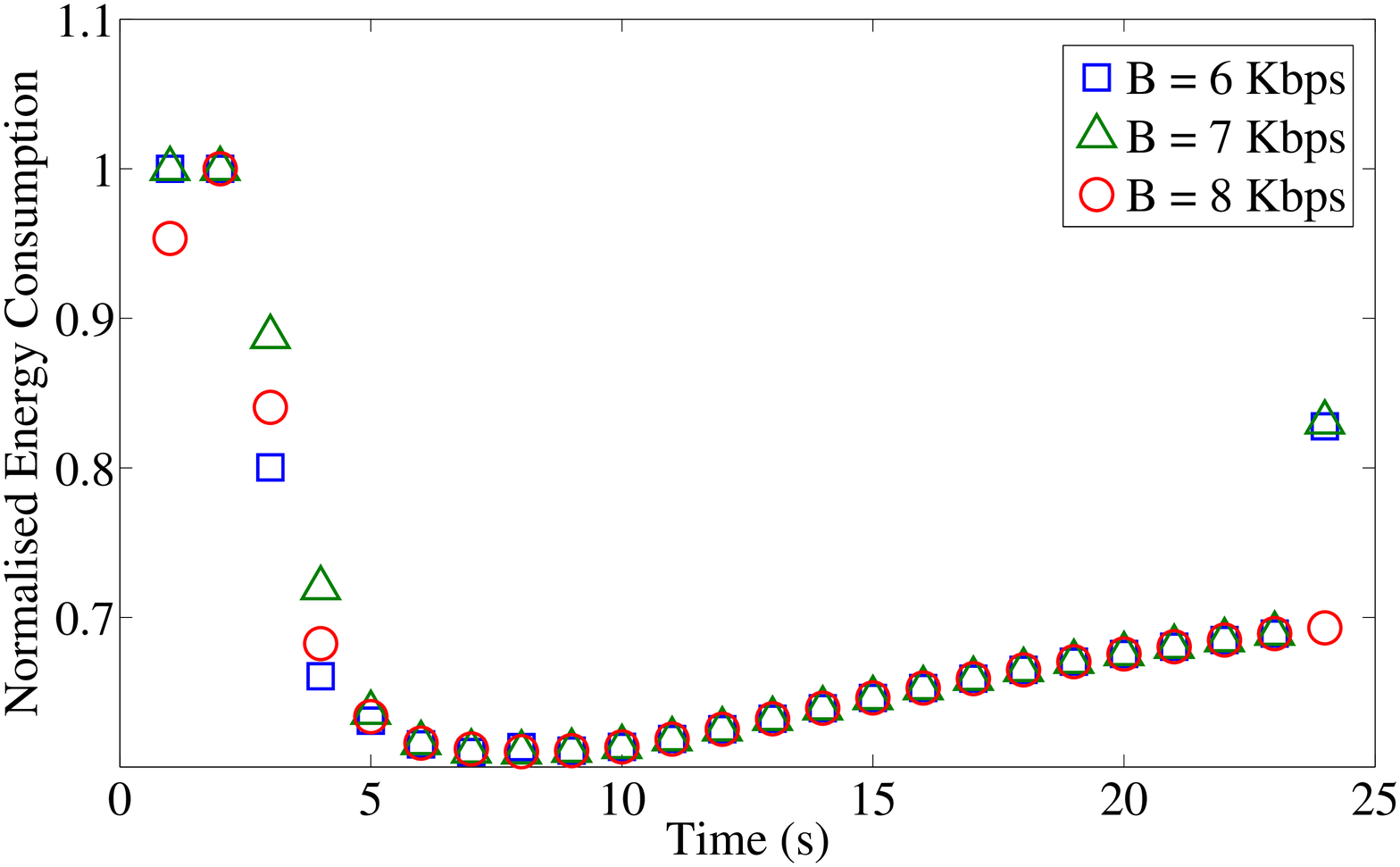}
\caption{Normalised total energy consumption for different channel bandwidths 
with respect to baseline scheme}
\label{fig:engobj}
\end{figure}

\begin{figure}[t]
\centering
\includegraphics[width=0.5\textwidth]{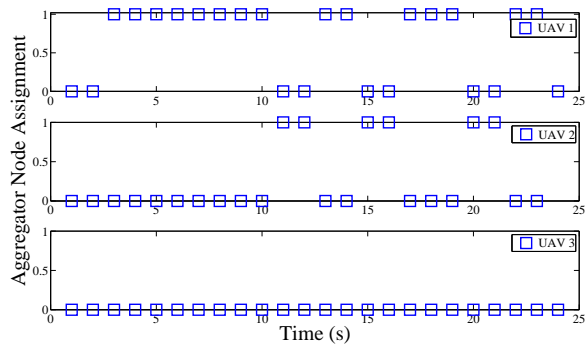}
\caption{Aggregator node assignments at different time steps for channel 
bandwidth $B=7$ Kbps}
\label{fig:AggAssgn}
\end{figure}

We compare the results obtained from the MINLP with a baseline strategy where all sensor nodes individually communicate with the base station using a single-hop communication protocol. The comparison is performed in terms of energy consumed per decision time interval $[t,t+h]$ between the MINLP and the baseline strategy. The vertical axis in Figure~\ref{fig:engobj} represents the system energy consumption per decision time $[t,t+h]$ normalized by the baseline scheme. Similar to the observations in~\cite{cotuk2014}, which studies the impact of bandwidth constraints of the energy consumption of WSN, our simulation also suggests that the channel bandwidth constraint has an effect on the energy consumption of the system. This is due to the restriction on the information flow pattern. Specifically, when the bandwidth is limited below the threshold value of 5 Kbps (not shown on the plot), the MINLP algorithm cannot find a solution that is better than the baseline strategy, hence no energy saving can be obtained. However, when the channel bandwidth is above the threshold, the MINLP can provide an optimal strategy that can save energy consumption up to 40\% compared to the baseline strategy, as shown in Figure~\ref{fig:engobj}. However, the energy saving improvement cannot be observed with an increase in $B > 6$~Kbps. Figure~\ref{fig:AggAssgn} shows the aggregator role assignments of each UAV at each time instance of the simulation, where 1 refers to an active role. 

\begin{figure}[t]
	\centering
	\includegraphics[width=0.4\textwidth]{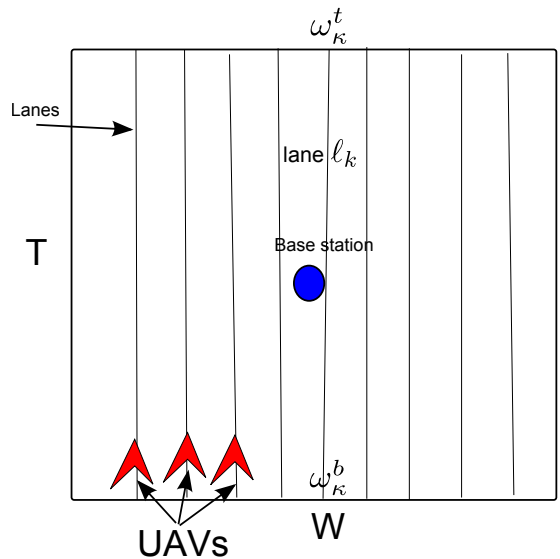}
	\caption{The search area is decomposed into lanes and each UAV is 
assigned to one lane. Once the UAV completes one lane, then  another lane is 
assigned.}\label{fig:scene}
\end{figure}

\subsection{Area mapping}\label{mapp}
A team of $n$ UAVs are deployed to survey a rectangular region with a length of $T$ meters and a width of $W$ meters using cameras. The vehicles are subject to communication, sensing and energy constraints.  Each UAV has a sensing range of 
$r_s$ meters determined by the camera resolution and altitude. Typically, mapping applications are performed using a lawn-mowing pattern and hence we split the rectangular region into lanes of  width $\zeta{r_s}$, where $0<\zeta\leq 1$ is the overlap factor. $\zeta=1$ implies the distance between the lanes is $r_s$ and there is no overlap of sensing regions between the aerial survey of UAVs, while $0<\zeta<1$ implies there is some overlap of the sensor footprint between two adjacent lanes. In terms of area coverage $\zeta=1$ is the best strategy. However, for mapping purposes, there must be at least $50\%$ overlap between two lanes to create good mosaics \cite{hudzietz2011experimental}. We assume a linear relationship between the overlap factor and the data aggregation ratio, i.e.~$\zeta = \gamma$, which means that the higher the overlapping area, the higher the data reduction after data processing. Note that here we assume that the overlap factor is a constant and the same for all nodes, therefore the subscript $z$ of $\gamma$ notation is dropped. The number of lanes are $N_{\ell}:= \lceil{{ \frac{T}{2\zeta r_s}}}\rceil+1$ and each lane is denoted by $\ell_{\kappa},\kappa=1,\ldots,N_{\ell}$. The vehicles use waypoint navigation for the survey and hence each lane $\ell_\kappa$ is represented by two waypoints $\ell_\kappa=(\omega_\kappa^b, {\omega_\kappa^t}),$ where, $\omega_\kappa^b=(x_\kappa^b,y_\kappa^b), \omega_\kappa^t=(x_\kappa^t,y_\kappa^t) $ as shown in Figure~\ref{fig:scene}. Lane $\ell_\kappa$ can be accurately tracked using any accurate path following 
algorithm \cite{sujit2014unmanned}. 

The time taken by the UAV team to survey the complete region depends on the number of UAVs deployed; when $n=1$, the lower bound on the mission time is  $WTN_\ell$ seconds.  Initially, UAV $i$ is given a lane $\ell_i,i\in N$ in terms of their waypoints $\ell_i=(\omega_i^b, {\omega_i^t})$. Once the vehicle reaches ${\omega_i^t}$, the lane  $\ell_{i+n}=(\omega_{i+n}^t,\omega_{i+n}^b)$ is assigned. However, we can see that UAV $i$ was assigned the waypoint sequence $(\omega_i^b, {\omega_i^t})$ for the first lane while  $(\omega_{i+n}^t,\omega_{i+n}^b)$ was assigned the next lane. If we assigned $(\omega_{i+n}^b,\omega_{i+n}^t)$, then the vehicle has to travel from $\omega_{i}^t$ to $\omega_{i+n}^b $, which is unproductive travel, 
since the vehicle expends fuel without surveying any of the region. Hence, we assign the UAV with an alternating sequence of waypoints. 

The desired heading angle $\psi_i^d$ is determined as 
\begin{eqnarray}
\psi_i^d = \left\{ 
\begin{array}{l l}
\arctan(y_{\kappa}^b-y_{\kappa}^t, x_{\kappa}^t-x_{\kappa}^b) & \quad \text{if 
$\ell_\kappa = (\omega_i^b, {\omega_i^t})$}\\
\arctan(y_{\kappa}^b-y_{\kappa}^t, x_{\kappa}^b-x_{\kappa}^t) & \quad \text{if 
$\ell_\kappa = (\omega_i^t, {\omega_i^b})$.}
\end{array} \right.
\end{eqnarray}

\subsubsection{Simulation Setting}
\begin{figure}
	\centering
	\includegraphics[width=0.45\textwidth]{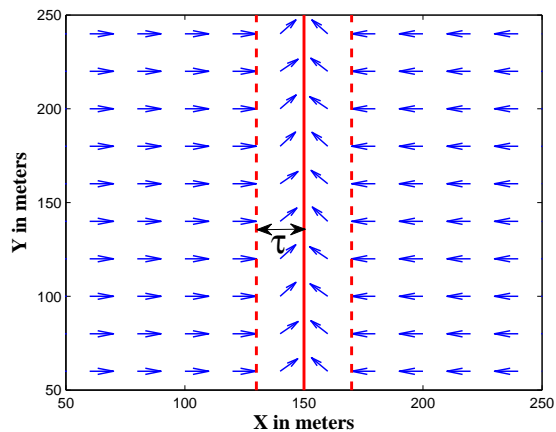}
	\caption{The path is given by waypoint (150,0) and (150,300). The vector field of the vehicle at various locations is shown. $\tau=20$ and $\chi=\pi/3$. }\label{fig:vectorfield}
\end{figure}

We consider a region of 3000\,m$\times$3000\,m and the base station is located in the middle at (1500,~1500). The sensing range of the vehicles $r_s=100$\,m, the communication range $r_c=500$\,m and the speed of the vehicles is~10\,m/s. We assume three vehicles are deployed to perform the mapping. The parameters used in the simulation are $\epsilon_s = 50$\,nJ/bit,~$\epsilon_p = 10$\,nJ/bit,~$\epsilon_r =135$\,nJ/bit,~$\epsilon_t=45$\,nJ/bit,~$\epsilon_{rf}=0.1$\,nJ/bit/m$^2$, $\beta = 2$, $L = 1280\times 720$ bits/packet and $\overline{\lambda}_z = 5$ packets/time interval. Each UAV communicates to the base station every $h=5$ seconds. The vector field based path following algorithm~\cite{nelson2007vector} is selected as the UAV path planning controller. The vector field based path following approach uses a two-fold strategy. When the vehicle is far away from the desired path, the algorithm directs the vehicle towards the path until the vehicle is~$\tau$ meters from the path as shown in Figure~\ref{fig:vectorfield}, where the parameter $\tau$ is the transition boundary between moving towards the path and following the path. The vehicle then transits into following the desired path with an entry angle of $\chi$. The effects of $\tau$ and $\chi$ are well studied in~\cite{nelson2007vector} and~\cite{sujit2012}. For all simulations, we use $\tau=20$ meters and $\chi=\pi/3$ rad. 

\begin{figure}[!h]
	\centering
\subfloat[]{\label{fig:z00}\includegraphics[width=4.5cm]{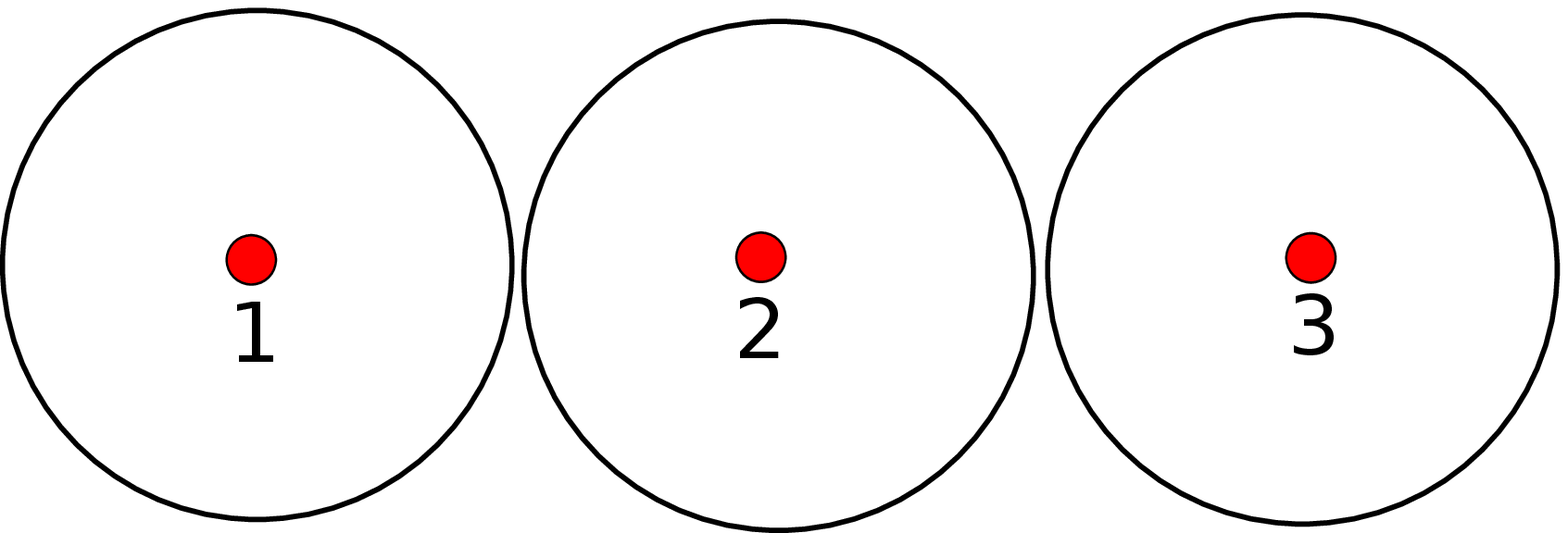}}\hspace{0.cm}
	\subfloat[]{\label{fig:z10}\includegraphics[width=4cm]{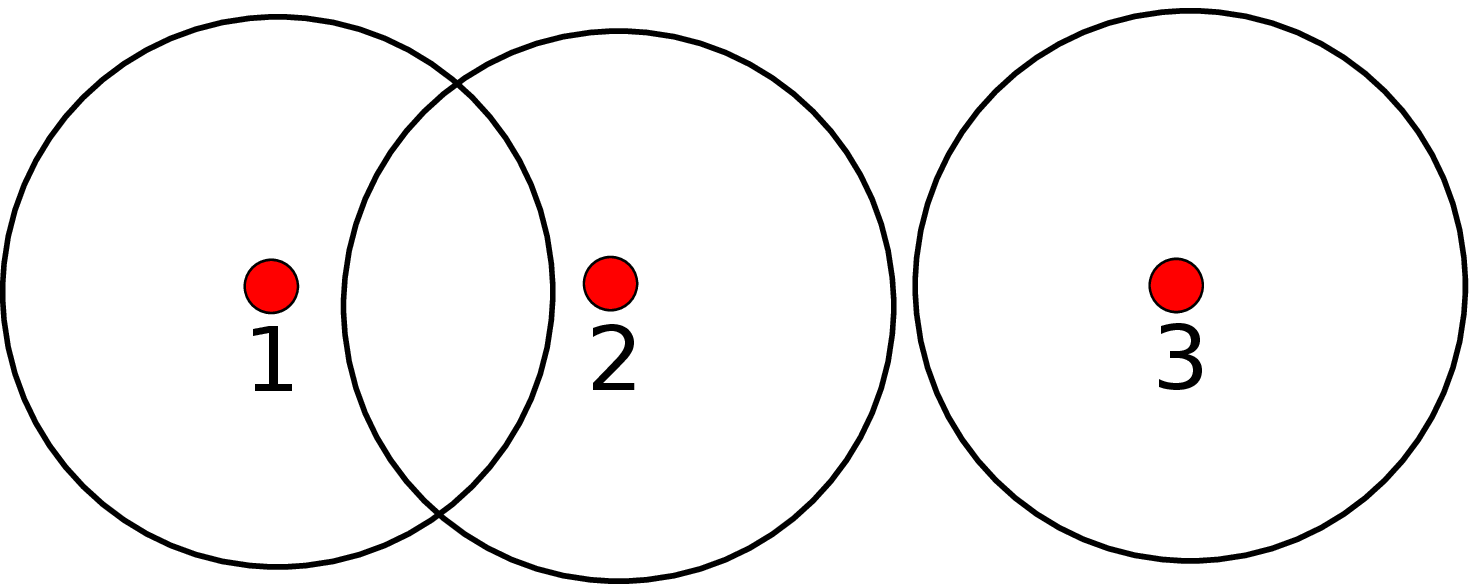}} 
		\caption{(a)  No common data between the nodes (b) nodes 1 and 2 have common data of type 0. }
	\end{figure}
	
\begin{figure}[h!]
\centering
\subfloat[]{\label{fig:z11}\includegraphics[width=3.5cm]{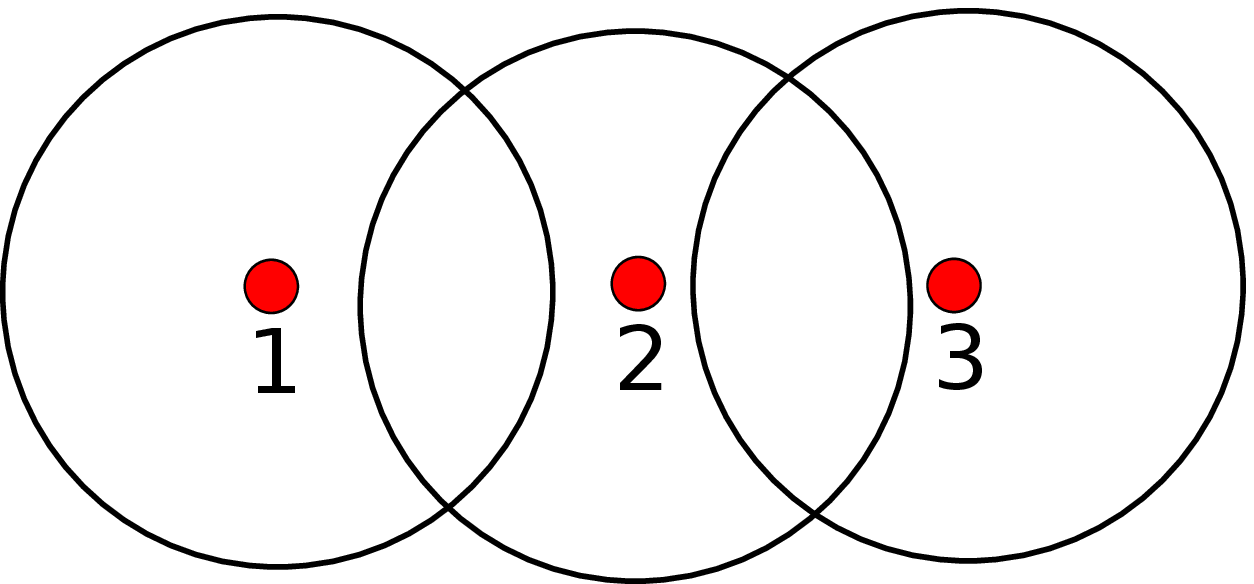}}\hspace{1cm}	
\subfloat[]{\label{fig:z111}\includegraphics[width=2.7cm]{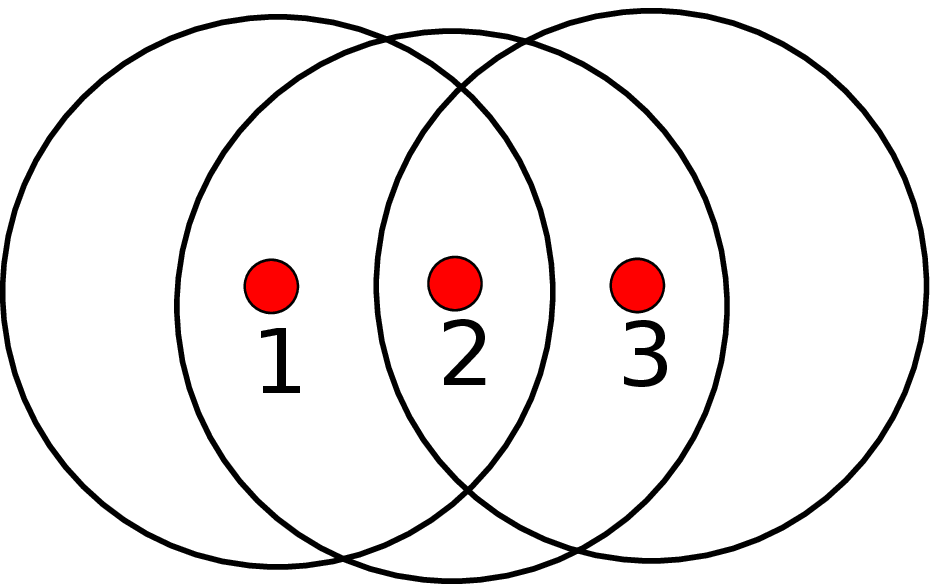}}
\caption{(a) node 2 has common data of type 0 with node 1 and type 1 with node 3 (b) all the nodes have common data of type 0 and 1.}
\end{figure}

For the mapping application, the values for $c_{0iz}$ depend on the distance between the nodes. That is, if the distance is greater than twice that of the sensing range $r_s$, then we will assume that the sensing data are not related and cannot be aggregated. In other words, the data are of different types. In order to illustrate how $c_{0iz}$ values are determined at each decision interval, consider a three vehicle system in Figure~\ref{fig:z00} where a distance between node $i$ and node $j$ $d_{ij}>2r_s$. For this scenario, there is no common data type between the nodes due to no overlap of the sensed 
regions, i.e.~$z\in\{0,1,2\}$. Therefore, the values of $c_{010}=1,c_{011}=0,c_{012}=0,c_{020}=0,c_{021}=1,c_{022}=0,c_{030}=0,c_{031}=0$ and $c_{032}=1$, which implies that none of the nodes have common data type.

Now consider the scenario as shown in Figure \ref{fig:z10}, where nodes 1 and 2 have a common data type $z=0$ and node 3 is distant from nodes 1 and 2. Therefore, in this case, we have $c_{010}=1,c_{011}=0, c_{020}=1,c_{021}=0, c_{030}=0$ and $c_{031}=1$. Similar to this scenario, if node $2$ and $3$ have a common data type $z=1$, while node 1 is distant from nodes $2$ and $3$, then $c_{010}=1,c_{011}=0, c_{020}=0,c_{021}=1, c_{030}=0$ and $c_{031}=1$. 

Another scenario is where one of the nodes may have two common data types, as shown in Figure~\ref{fig:z11} in which node $2$ shares data with node $1$ and node $3$, but node $1$ and node $3$ are far from each other and do not have common data. In this case, we set $c_{010}=1,c_{011}=0, c_{020}=1,c_{021}=1, c_{030}=0$ and $c_{031}=1$. The last scenario is where all nodes  are within $2r_s$ distance of each other as shown in Figure~\ref{fig:z111}. In this case, $c_{010}=1, c_{020}=1$ and $c_{030}=1$. Thus, depending on the node positions and overlap regions, the values $c_{0iz}$ are pre-determined at the beginning 
of each decision interval. For our simulations, we consider scenario as in Figure~\ref{fig:z11} for $\zeta<0.75$, and Figure \ref{fig:z111} for $\zeta>0.75$.

\subsubsection{Simulation Results}
\begin{figure}
	\centering
	\includegraphics[clip,trim=80 5 50 5,width=8cm,height=5cm]{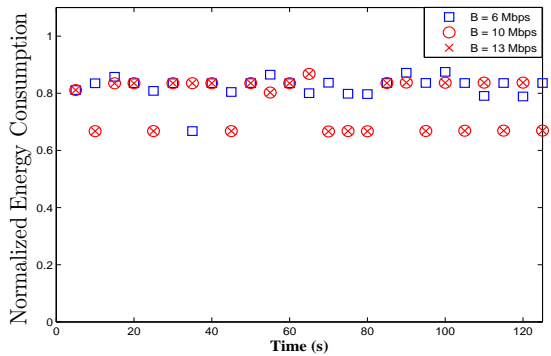}
	\caption{The normalized total energy of the MINLP compared to the baseline strategy for different bandwidth constraints having $\zeta=0.5$.}\label{fig:mappingbwagents}
\end{figure}

The bandwidth allocated to communicate with the base station plays a key role in determining the computing nodes. Figure \ref{fig:mappingbwagents} shows the total energy consumption of the MINLP normalised to the baseline strategy for every $h=5$ seconds with an overlap factor $\zeta=0.5$. When the available bandwidth is less than 6 Mbps (not shown on the plot), the nodes communicate directly to the base station. Hence, we do not show this effect. However, when we increase the bandwidth, data aggregation behaviours can be observed. As shown in  Figure~\ref{fig:mappingbwagents}, the energy saving is close to 20\% for most of the decision cycles (for $B=6$ Mbps). With further increase in bandwidth to $B=10$ Mbps, we can see that there is further increase in energy saving of 35\%. {\color{black}However, with additional increase in bandwidth to $B=13$ Mbps, there is no further improvement in energy saving.} As expected, the energy reduction is due to co-operation among the agents, i.e.~when the bandwidth is sufficiently large, more energy-efficient feasible information flow patterns are allowed.

\begin{figure}
	\centering
	\subfloat[B = 6 Mbps]{\includegraphics[width=0.5\textwidth]{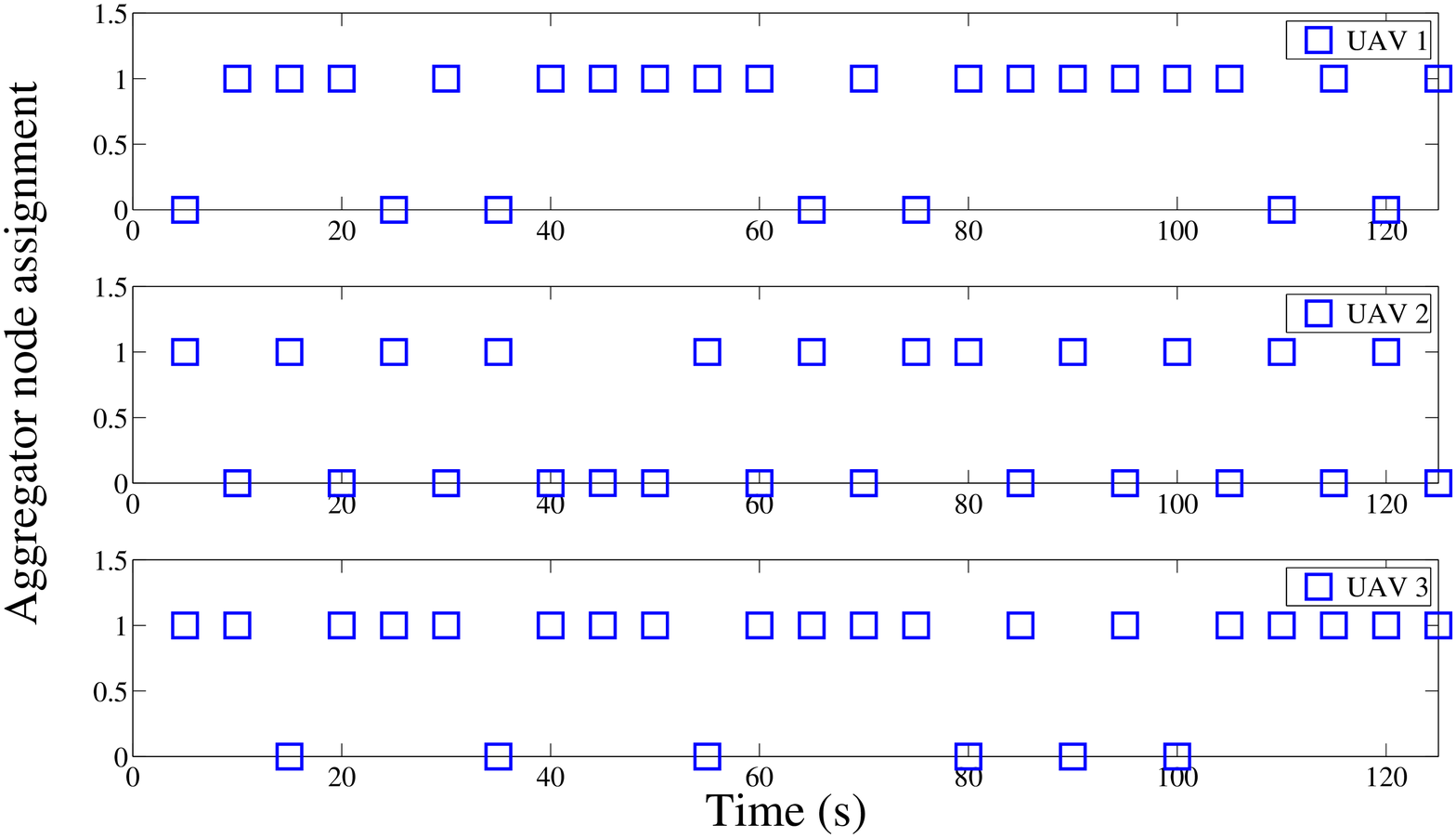}}
	\subfloat[B = 10 Mbps]{\includegraphics[width=0.5\textwidth]{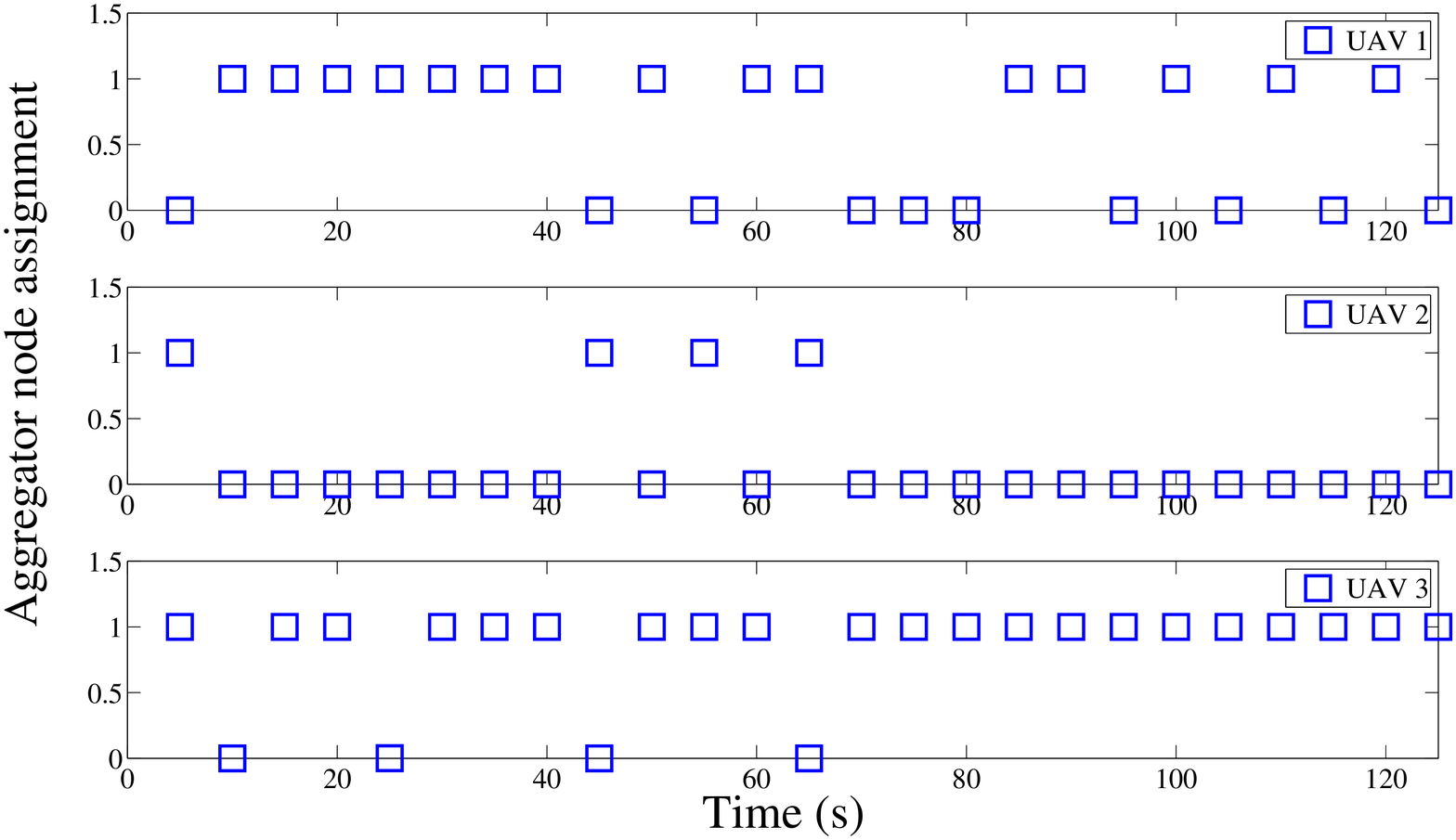}}
	\caption{The aggregator node selection  at different time steps when the bandwidth parameter is varied for the same overlap $\zeta=0.5$.} 
\label{fig:mappingaggagents}
\end{figure}

\begin{figure}
	\centering
	\includegraphics[clip,trim=80 5 5 5,width=0.5\textwidth]{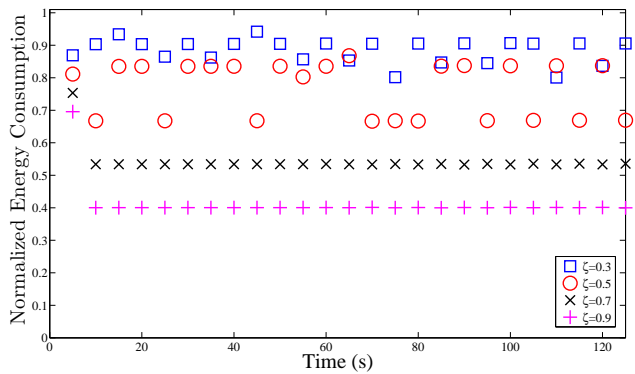}
	\caption{The normalized total energy of the MINLP with reference to the baseline strategy for different overlap factors having a channel bandwidth of $B=10$ Mbps.} \label{fig:zetaEffect}
\end{figure}

In the mapping application, the overlap factor $\zeta$ plays a key role in determining the amount of information that needs to be transmitted by the aggregator node to the base station. When $\zeta$ increases, the agents are close to each other with high overlap. Therefore, during the mosaic operation, the resultant image size will be smaller compared to the sum of individual images. In order to validate this hypothesis, we carried out experiments with different overlap factors $\zeta=0.3,0.5,0.7$ and $0.9$ for the same bandwidth of 10 Mbps. In Figure~\ref{fig:zetaEffect}, we can see the effect of $\zeta$ for 
a given bandwidth. Specifically, the energy saving increases as $\zeta$ increases. For example, when $\zeta=0.9$, we can achieve savings up to 60\% compared to the baseline strategy. 

{\color{black}
\begin{figure}
	\centering
	\includegraphics[clip,trim=120 5 40 5,width=0.5\textwidth,height=6cm]{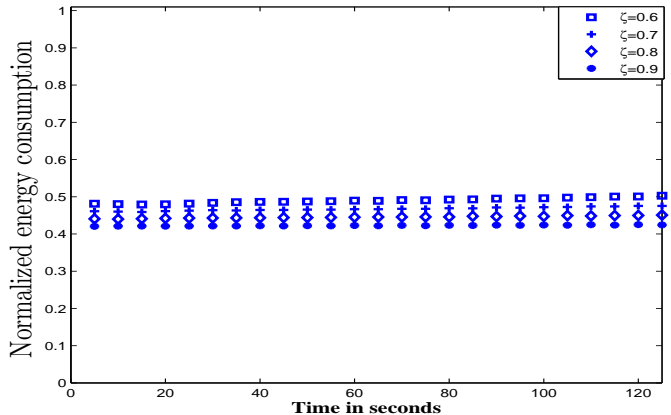}
	\caption{The normalized total energy of the MINLP (for 5 agents) with reference to the baseline strategy for different overlap factors having a channel bandwidth of $B=20$ Mbps.} \label{fig:a5-nf}
\end{figure}
We further, carry out simulations for 5 agents having the same simulation parameters as above. Figure \ref{fig:a5-nf} shows the respective energy saving when 5 agents perform the survey. With increasing overlap factor, the amount of information to be dispatched reduces and hence there is  a decrease in energy consumption. With increase in number of agents we can see that a trend in energy conversation similar to that of agent 3 simulation can be seen. 

We are not performing simulations with large number of agents ($>$10) because (i) normally, UAVs are composed of no more than 10 in a real application, which is different from WSN which are composed of a large number of nodes and (ii) if we were to apply our approaches to a large number of UAVs (10+), then we can adopt a hierarchical approach, where a small set of UAVs ($
\leq 10$) are assigned to a single base-station and the operation consists of many base stations. With increase in number of nodes, the amount of data to be transmitted increases and a single receiver may not be able to handle such high traffic. Hence, the usual approach especially when imagery data need to be transmitted from UAVs is to assign a receiver to which a small set of UAVs communicate.
}
\section{Conclusions}\label{sec:conclusions}
Cooperation between mobile computing agents enables them to optimize the computation and communication energy consumption, thereby increasing the system lifetime. We have devised an MINLP formulation that shows lower energy consumption by incorporating data aggregation and clustering schemes. The MINLP formulation is generic and we utilized this generality by validation on two data gathering applications, namely target tracking and mapping. We have studied the effect of different parameters on the MINLP decision-making. Simulation results show that the channel bandwidth has a direct impact on the energy saving scheme, i.e. sufficient bandwidth is necessary for an implementation of an intelligent information routing scheme.    

The proposed MINLP formulation can be further extended to optimize the energy consumption of various units. One potential direction is to make a decision on when to communicate to the base station. Currently, we assume that the decision interval is fixed. However, depending on the amount of data, channel bandwidth and the transceiver energy properties, the decision cycle can be dynamically selected to optimize the overall energy consumption.

Solving the MINLP efficiently as well as whether to implement the proposed framework in a centralized or distributed manner could be subjects for future work.

\bibliographystyle{IEEEtran}
\bibliography{uavtracking}

\end{document}